\documentstyle[epsf,psfig]{article}
\def\:{\hbox{\bf :}}\def\d{\mbox{d}}

\def\thesection{\arabic{section}}
\newcommand{\sect}[1]{\stepcounter{section}\section*{\protect{
\noindent\large\bf\thesection~\normalsize\bf #1}}}

\def\thefigure{\arabic{figure}}
\def\fnum@figure{{\bf Figure \thefigure}}
\begin{document}
\def\pni{\par\noindent}
\pagestyle{myheadings}
\markboth{\qquad G. M. D'Ariano, M. Rubin, M. F. Sacchi 
and Y. Shih\hspace*{\fill}}
{\hspace*{\fill}
Quantum tomography of the GHZ state\qquad}
\vspace*{1mm}
\begin{center}
{\Large\bf Quantum tomography of the GHZ state}
\end{center}
\begin{center}
by {\large{\sl G. M. D'Ariano$^{a}$, M. Rubin$^{b}$, 
M. F. Sacchi$^{a}$, and Y. Shih$^{b}$}}, 
\end{center}
\begin{center}
$^a$ Dipartimento di Fisica \lq A. Volta\rq, 
Universit\`a di Pavia and INFM Unit\`a di Pavia\\ 
via A. Bassi 6, I-27100 Pavia, Italy\\
$^b$ Physics Department, University of Maryland, Baltimore County,\\
Baltimore, Maryland 21228
\end{center}
{
\begin{quote}{\bf Abstract.}
We present a method of generation of the 
Greenberger--Horne--Zei\-lin\-ger state involving type II and type I
parametric downconversion, and triggering photodetectors. The state
generated by the proposed experimental set-up can be
reconstructed through multi-mode quantum homodyne tomography.  The
feasibility of the measurement is studied on the basis of Monte-Carlo
simulations.
\end{quote}}
\sect{Introduction} A number of proposals for generating the
Greenberger--Horne--Zeilinger (GHZ) state \cite{orig} has been
suggested in the literature \cite{lit}. Such kind of state is very
interesting as it leads to correlations between three particles in
contradiction with the Einstein-Podolsky-Rosen idea of ``elements of
reality'' \cite{epr}. In the present contribution we present a scheme
for a complete quantum test of a GHZ state of radiation, not just for
a simple verification of some GHZ correlations, which do not prove
that a true GHZ state has been produced.  In fact, the verification of
a state-preparation procedure needs a complete state-reconstruction
technique, whereas correlation measurements \cite{zeil} 
give identical results for 
very different states of radiation. In this respect, a crucial
technique for state-preparation tests is quantum homodyne tomography,
in which the detrimental effect of non-unity quantum efficiency of
detectors is taken into account {\em ab initio} by the reconstruction
algorithms.  \par
\begin{figure}[hbt]
\begin{center}\epsfxsize=.65 \hsize\leavevmode\epsffile{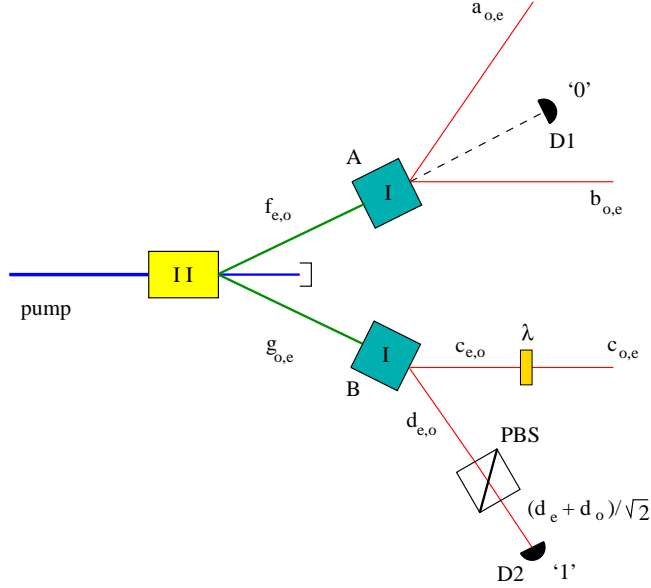}
\end{center}
\caption{\scriptsize Sketch of the experimental set-up for generating the GHZ
state. 
A low-gain type-II parametric downconverter is pumped by a
strong coherent beam and generates the state in Eq. (\ref{0ebell}). 
Two further downconversion processes in type I nonlinear crystals 
excite modes $a_{o,e},\,b_{o,e},\,c_{o,e}$ and $d_{o,e}$.  
$\lambda $ and PBS denote a wave plate and a polarizing beam splitter,
respectively. The photodetector in the lower part of the scheme
collapses the field modes in the state of Eq. (\ref{fi}). The 
photodetector in the upper part checks if the second photon in one of
the modes $f_{e,o}$ has been split in the first crystal. The resulting
state is a mixture of a Fock state in modes $c_{o,e}$ and of a GHZ 
state [see Eq. (\ref{mixt})], 
the weights of the two components depending on the effective
gain of crystal A and on the quantum efficiency of detector D1.}
\label{f:experiment}\end{figure}

 In the following we propose a method for generating
a GHZ state through type II and type I parametric downconversion, and
triggering photodetectors. The proposed set-up, although it has low
rate of production due to low efficiency for single-photon
downconversion, however is the only way to generate a ``true'' GHZ
state, without an additional vacuum component. 
The scheme allows a tomographic state-reconstruction, whose
feasibility here is studied on the basis of
Monte--Carlo simulations. 
\sect{Scheme for the GHZ-state generation} The scheme
for the generation of the GHZ state is sketched in
Fig. \ref{f:experiment}.  A low-gain type-II parametric downconverter
is pumped by a strong coherent beam to generate the state
\begin{eqnarray}
|\xi \rangle \simeq (1+2\gamma ^2)^{-1/2}\left[
|0 \rangle + \gamma \left( |1f_e,1g_o \rangle +e^{i\varphi _1}
|1f_o,1g_e \rangle  \right) 
\right] \;,\label{0ebell}
\end{eqnarray}
in the four radiation modes $f_{e,o}$ and $g_{o,e}$, where $o,e$
represent ordinary and extraordinary polarizations, $\gamma $ denotes
the effective coupling depending on the pump strength and the
nonlinear susceptibility of the crystal, $|0\rangle $ and $|1a\rangle$
represent the vacuum and the single-photon Fock state for mode $a$,
respectively.
\par
\begin{figure}[ht]
\centerline{\psfig{file=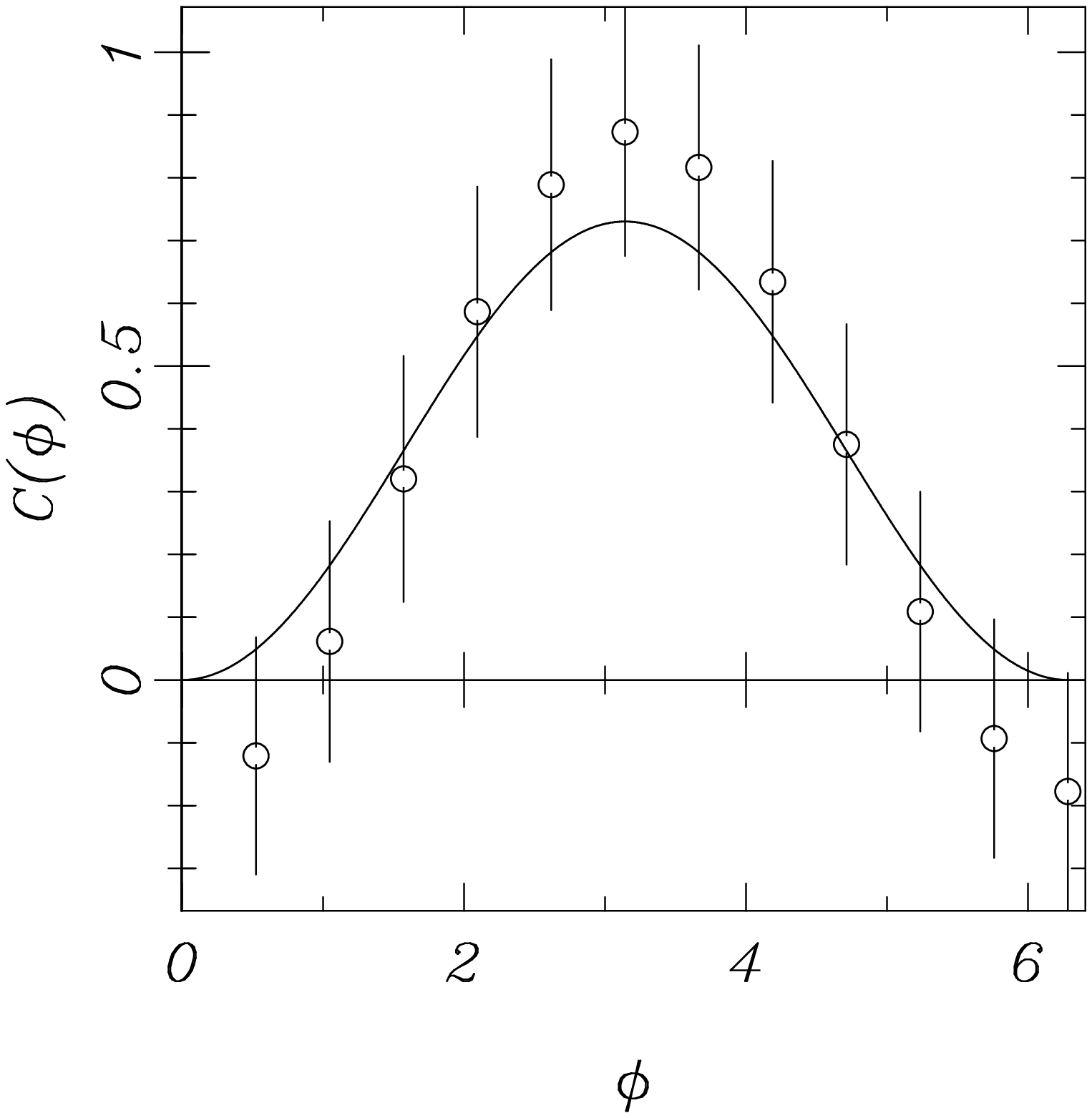,width=5.cm}
\hspace{60pt}
\psfig{file=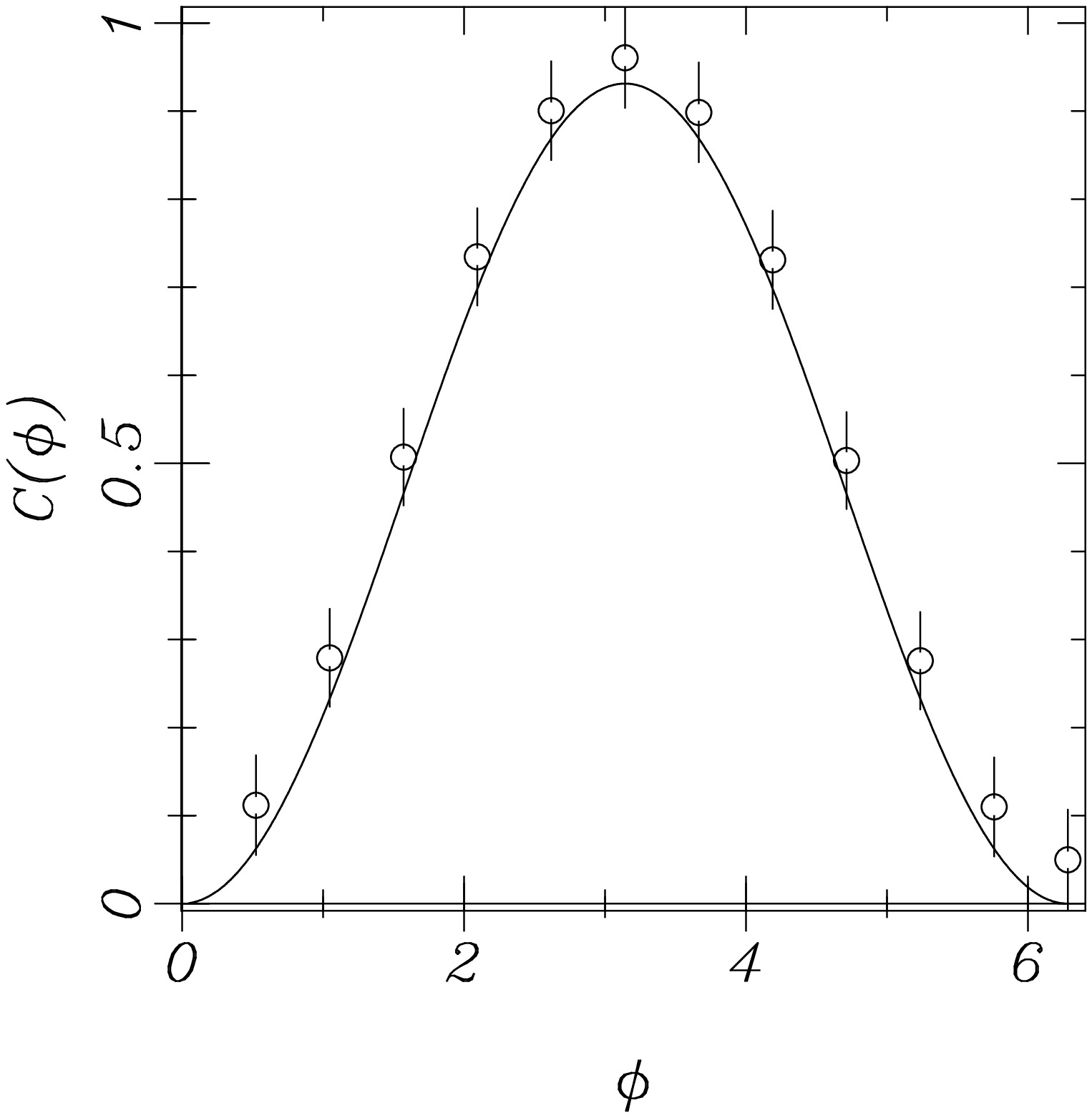,width=5.cm}} 
\caption{\scriptsize 
Tomographic measurement of $C(\phi)$ in Eq. (\ref{cfi}) for
$\varphi =0$. On the left: 
$\eta =0.85$, $\chi =0.3\,\pi$, $N=6\times 10^6$. On the right:  
$\eta =0.9$, $\chi =0.4\,\pi$, $N=1.7\times 10^7$.} 
\label{tomoghz}\end{figure}

 The state $\xi $ at the output of the first crystal is then
impinged on two type-I nonlinear crystal A and B. In this case 
no classical pump is used and the dynamics must be evaluated without 
the parametric approximation used to derive Eq. (\ref{0ebell}). 
For example, the unitary evolution describing crystal A is given by
\begin{eqnarray}
\hat U_A =\exp{\left[\chi_A \left(a_o^{\dag }b_o^{\dag }f_e +
e^{i\varphi _A}\,a_e^{\dag }b_e^{\dag }f_o -\hbox{h.c.}
 \right)\right]}\;, 
\end{eqnarray}
$\chi_A$ being proportional to the nonlinear susceptibility of the
medium. An analogous expression can be written for the crystal B. For
simplicity, we assume in the following $\chi _A=\chi _B\equiv \chi$. 
The state at the output of the couple of type-I crystals is given by 
\begin{eqnarray}
&&\hat U_A\hat U_B |\xi \rangle =
(1+2\gamma ^2)^{-1/2} \{|0 \rangle +\nonumber \\  &&
\gamma [(\cos\chi |1f_e \rangle +\sin\chi
|1a_o,1b_o \rangle ) (\cos\chi |1g_o \rangle +\sin\chi |1c_e\,1d_e
\rangle )+\nonumber \\&&
e^{i\varphi_1} 
(\cos\chi |1f_o \rangle +e^{i\varphi _A}\sin\chi
|1a_e,1b_e \rangle ) (\cos\chi |1g_e \rangle +e^{i\varphi_B}
\sin\chi |1c_o\,1d_o )\rangle ]\}\;. 
\end{eqnarray}
As shown in Fig. \ref{f:experiment}, a wave plate and a polarizing
beam splitter act respectively on modes 
$c_{e,o}$ and $d_{e,o}$ according to the unitary transformations
\begin{eqnarray}
\left\{\begin{array}{l}
c_e \longrightarrow \ c_o \\
c_o \longrightarrow -c_e \end{array}\right.
\;,
\qquad \quad\left\{\begin{array}{l}
d_e \longrightarrow (d_e+d_o)/\sqrt 2 \\
d_o \longrightarrow (d_o-d_e)/\sqrt 2 \end{array}\right.
\;. 
\end{eqnarray}
When photodetector D2 reveals one photon after the polarizing beam 
splitter, one is guaranteed that two photons have been created in the
type-II crystal and that the photon impinging on crystal B has been
split. This occurs with probability 
$P_{\Phi}=\eta_2(\gamma ^2\sin ^2\chi)/(1+2\gamma ^2)$, 
$\eta _2$ being the quantum efficiency of photodetector D2. 
The corresponding reduced state $|\Phi \rangle  $ writes
\begin{eqnarray}
|\Phi \rangle &=&\frac {1}{\sqrt 2}[
\cos\chi (|1f_e\,1c_o \rangle -e^{i(\varphi _1 +\varphi _B)}|1f_o\,1c_e
\rangle )
\nonumber \\&+& 
\sin\chi (|1a_o\,1b_o\,1c_o \rangle -e^{i(\varphi _1 +\varphi_A +\varphi _B)}
|1a_e\,1b_e\,1c_e \rangle )]
\;.\label{fi}
\end{eqnarray}
Photodetector D1 in the upper part of the scheme monitors the splitting of the photon
impinging on crystal A. When such detector, characterized by quantum
efficiency $\eta _1$, reveals the lack of the ``pumping'' photon, the
resulting output state finally reads as the following mixture
\begin{eqnarray}
\hat\varrho =\left\{\begin{array}{ll}\; 
\frac{1}{\sqrt 2}
(|1a_o\,1b_o\,1c_o \rangle -e^{i(\varphi _1 +\varphi_A +\varphi _B)}
|1a_e\,1b_e\,1c_e \rangle ) & \ \ p_1=\frac {\sin^2\chi}{1-\eta _1\cos
^2\chi}\\
|1c_o \rangle & \ \ p_2=\frac {(1-\eta _1)\cos^2\chi}{2(1-\eta _1\cos
^2\chi)}\\
|1c_e \rangle & \ \ p_3=p_2 \end{array}\right.
\;\label{mixt}
\end{eqnarray}
The overall probability $P_{\varrho}$ 
of generating the mixture in Eq. (\ref{mixt}) 
is given by $P_{\varrho}= P_{\Phi }(1-\eta _1\cos^2\chi)$.
One easily recognizes in the first component of the mixed state in
Eq. (\ref{mixt}) the GHZ state for a suitable arrangement of the
phases, namely for $\varphi _1 +\varphi_A +\varphi _B=\pi $.
\sect{Multi-mode tomographic measurement}\label{due} 
Quantum homodyne
tomography is the first quantitative technique for measuring the
matrix elements of the radiation density operator \cite{pra50,bilk},
which is now used in optical labs \cite{breiten}.  Single-mode
homodyne tomography can be generalized to any number of modes. 
However, such a simple generalization needs a separate measurement for each
mode, which cannot be achieved when modes are not spatially
separated. For this reason, in Ref. \cite{onelo} it has been proposed  
a general method for measuring an arbitrary 
observable of a multi-mode electromagnetic field, using homodyne
detection with a {\em single} local oscillator. Such method is a
natural application of a recent group-theoretical approach to quantum
tomography \cite{acta}. In the following we recall the main results, 
providing  the rule 
to evaluate the ``unbiased estimator'' for a generic (M+1)-mode
operator. The quantum expectation value of the operator can be
obtained for any unknown state of radiation through an average of such
estimator over homodyne outcomes that are collected using a single
local oscillator by scanning different linear combinations of modes on
it. The quadrature operator to be measured is given by 
$\hat X({\bf\theta},{\bf\psi})=[\hat
A^{\dag}({\bf\theta},{\bf\psi})+\hat
A({\bf\theta},{\bf\psi})]/2$ with $\hat
A({\bf\theta},{\bf\psi})= \sum_{l=0}^M
e^{-i\psi_l}u_l({\bf\theta})a_l$. 
The vector $\vec u(\theta)$ represents a point on the Poincar\'e
hyper-sphere (for the explicit parametrization see
Ref. \cite{onelo}).  By scanning the values of $\psi_l\in[0,\pi]$ and
$\theta_l\in[0,\pi/2]$, all possible linear combinations of modes
described by annihilation operators $a_l$, with $l=0,\ldots,M$, are
obtained.  The homodyne outcomes for $\hat X({\bf\theta},{\bf\psi})$
can be obtained using a single local oscillator prepared in the
multi-mode coherent state $\otimes _{l=0}^M |\gamma _l \rangle $ with
$|\gamma _l \rangle =e^{i\psi _l}u_l(\theta) K/2$, where $K \gg
1$. The expectation value for a given operator $\hat O$ is evaluated
as follows
\begin{eqnarray}
\langle\hat O\rangle=\int\d\mu[{\bf\psi}]\int\d\mu[{\bf\theta}]\,
\int _{-\infty}^{+\infty}\d x \,p_\eta (x;{\bf\theta},{\bf\psi})
\,{\cal E}_{\eta}[\hat O](x;{\bf\theta},{\bf\psi})\;,\label{K}
\end{eqnarray}
where $p_{\eta}(x;{\bf\theta},{\bf\psi})$ denotes the homodyne
probability distribution of the quadrature $\hat
X({\bf\theta},{\bf\psi})$ for quantum efficiency $\eta$, and 
the function ${\cal E}
_{\eta}[\hat O](x;{\bf\theta},{\bf\psi})$ of $x$, $\{\theta_l\}$,
$\{\psi_l\}$ has the following analytic expression
\begin{eqnarray}
{\cal E}_{\eta}[\hat O](x;{\bf\theta},{\bf\psi})=\frac{\kappa^{M+1}}{M!}
\int_0^{+\infty} \d t\, e^{-t+2i\sqrt{\kappa t}x}\, t^M\, \mbox{Tr}\{\hat
O\:\exp[-2i\sqrt{\kappa t}\hat X({\bf\theta},{\bf\psi})]\:\}
\;,\label{MAIN}
\end{eqnarray}
with $\kappa=2\eta/(2\eta-1)$. In Eq. (\ref{MAIN}) we used the notation
\begin{eqnarray}
\int\d\mu[{\bf\psi}]\doteq\prod_{l=0}^M\int_0^{2\pi}
\frac{\d\psi_l}{2\pi}\;,\qquad
\int\d\mu[{\bf\theta}]
\doteq 2^M \,M!\prod_{l=1}^M\int_0^{\pi/2}d\theta_l\,
\sin^{2(M-l)+1}\theta_l\cos\theta_l\;.
\label{notaz}
\end{eqnarray}
For any
given operator $\hat O$ Eq. (\ref{MAIN}) provides the ``unbiased
estimator'' to be averaged over all homodyne outcomes for the
quadrature $\hat X({\bf\theta},{\bf\psi})$ of all modes in order to
obtain the ensemble average $\langle\hat O\rangle$ for any unknown
state of radiation. Eq. (\ref{K}) can be specialized to the matrix element
$\langle\{n_l\}|\hat R|\{m_l\}\rangle$ of the full joint density matrix. This
will be obtained by averaging the following estimator \cite{onelo}
\begin{eqnarray}
&&{\cal E}_{\eta}[|\{m_l\}\rangle\langle\{n_l\}|](x;{\bf\theta},{\bf\psi})=
e^{-i\sum_{l=0}^M(n_l-m_l)\psi_l}\,
\frac{\kappa^{M+1}}{M!}
\prod_{l=0}^M\left\{[-i\sqrt{\kappa} u_l({\bf\theta})]^{\mu_l-\nu_l}
\sqrt{\frac{\nu_l!}{\mu_l!}}\right\}\nonumber\\ &&\times
\int_0^{+\infty}\d t\,e^{-t+2i\sqrt{\kappa t}x}\,
t^{M+\sum_{l=0}^M(\mu_l-\nu_l)/2}\prod_{l=0}^ML_{\nu_l}^{\mu_l-\nu_l}
[\kappa u_l^2({\bf\theta})t]\;,\label{gasp}
\end{eqnarray}
where $\mu_l=\mbox{max}(m_l,n_l)$, $\nu_l=\mbox{min}(m_l,n_l)$, and
$L_n^{\alpha}(z)$ denotes the customary generalized Laguerre
polynomial of variable $z$. 
\sect{Numerical results}
The tomographic measurement on the state in
Eq. (\ref{mixt}) can be suitably performed by varying randomly the 
phases and the polarizations  
for the couples of modes $a_{o,e}$, $b_{o,e}$ and $c_{o,e}$ and then 
collecting homodyne outcomes by using three 
local oscillators. Such an experimental arrangement represents an
intermediate way of using the multi-mode tomographic method of
Sect. 3, and the usual method based on the product of
single-mode estimators. By using the estimator in Eq. (\ref{gasp}) 
one can measure tomographically the expectation value of the
projector $|\phi \rangle \langle \phi |$ with
\begin{eqnarray}
|\phi \rangle \equiv \frac {1}{\sqrt 2}\left(
|1a_o\,1b_o\,1c_o \rangle +e^{i\phi }
|1a_e\,1b_e\,1c_e \rangle \right)
\;\nonumber \label{proj}
\end{eqnarray}
on the state
$\hat\varrho $ in Eq. (\ref{mixt}) 
and compare the result
with the theoretical value, namely
\begin{eqnarray}
C(\phi )\equiv\langle \phi |\hat\varrho|\phi \rangle =\frac 12
\,p_{1}\,\left[1-\cos(\phi-\varphi)\right]\;\label{cfi}
\end{eqnarray}
We report in Fig. \ref{tomoghz} the
results of some Monte-Carlo simulations of the tomographic measurement
of $C(\phi )$ in Eq. (\ref{cfi}). 
\par\noindent In the simulations the quantum
efficiency of detectors D1 and D2 is $\eta _1=\eta _2=30\%$ and the
phase $\varphi $ in the state (\ref{mixt}) is $\varphi =0$.  The
values of the quantum efficiency $\eta $ of homodyne detectors, 
the coupling $\chi $
of type-I downconvertors, and the number $N$ of simulated homodyne
data are reported in the caption of the figures.  
The results of the simulations show that for homodyne detectors 
with quantum efficiency $\eta =85\%$ one needs about $10^7$ data to 
reconstruct the state with relatively small statistical error. 
The experimental values
compare very well with the theoretical ones.  The bars represent the
statistical error, whereas the solid line is the theoretical value of
$C(\phi)$.  In each plot, all points are obtained by the same sample 
of data which 
causes the evident correlation between the statistical deviations.
\section*{\normalsize\bf References}
\begin{description}
\bibitem[1]{orig} D. M. Greenberger, M. Horne, and A. Zeilinger, in {\em
Bell's Theorem, Quaantum Thoery, and Conceptions of the Universe},
M. Kafatos, Ed. (Kluwer, Dordrecht 1989) pp. 69-72. 
\bibitem[2]{lit} See T. E. Keller, M. H. Rubin, Y. H. Shih, and L. Wu, 
Phys. Rev. A {\bf 57}, 2076 (1998), and references therein.
\bibitem[3]{epr} A. Einstein, B. Podolsky, and N. Rosen, Phys. Rev.  {\bf
47}, 777 (1935).
\bibitem[4]{zeil} D. Bouwmeester, J. Pan, M. Daniell, H. Weinfurter,
and A. Zeilinger, Phys. Rev. Lett. {\bf 82}, 1345 (1999). 
\bibitem[5]{pra50} G. M. D'Ariano, C. Macchiavello, and M. G. A. Paris, 
Phys. Rev. A {\bf 50}, 4298 (1994).
\bibitem[6]{bilk} G. M. D'Ariano, 
in {\em Quantum Optics and the Spectroscopy of Solids}, 
T. Hakio\v{g}lu and A. S. Shumovsky, Eds. (Kluwer, Dordrecht 1997) 
pp. 139-174. 
\bibitem[7]{breiten} G. Breitenbach, S. Schiller and J. Mlynek, 
Nature {\bf 387}, 471 (1997).
\bibitem[8]{onelo} G. M. D'Ariano, P. Kumar, and M. F. Sacchi, unpublished.
\bibitem[9]{acta} G. M. D'Ariano, Acta Physica Slovaca {\bf 49}, 513
(1999).
\end{description}
\end{document}